\documentstyle[12pt]{article}
\def\le{\langle}
\def\re{\rangle}

\def\1{\mbox{I\hspace{-.15em}1}}

\def\re{\rangle}

\def\b{\begin{equation}}
\def\e{\end{equation}}
\def\bee{\begin{enumerate}}
\def\eee{\end{enumerate}}

\title{A Naturally Renormalized Quantum Field Theory}
\author{S. Rouhani$^1$ and M.V. Takook$^{2}$\thanks{e-mail:
takook@razi.ac.ir}}

\begin{document}
\maketitle \centerline{\it $^1$Plasma Physics Research Center,
Islamic Azad University,} \centerline{\it P.O.BOX 14835-157, Tehran,
IRAN}\centerline{\it $^2$Department of Physics, Razi University,
Kermanshah, IRAN}

\begin{abstract}

It was shown that quantum metric fluctuations smear out the
singularities of Green's functions on the light cone \cite{for2},
but it does not remove other ultraviolet divergences of quantum
field theory. We have proved that the quantum field theory in
Krein space, {\it i.e.} indefinite metric quantization, removes
all divergences of quantum field theory with exception of the
light cone singularity \cite{gareta,taijmpe1}. In this paper, it
is discussed that the combination of quantum field theory in Krein
space together with consideration of quantum metric fluctuations,
results in quantum field theory without any divergences.

\end{abstract}

\vspace{0.5cm} {\it Proposed PACS numbers}: 04.62.+v, 03.70+k,
11.10.Cd, 98.80.H \vspace{0.5cm}

\section{Introduction}

One of the greatest challenges of physics today is the achievement
of a proper theory of quantized gravitational fields. In other words
a theory, which quantizes the gravitational fields without any
anomaly, is sought for the past seven decades without a thorough
success. The element of time, which has two completely different
concepts in quantum mechanics as opposed to general relativity, is
the most outstanding problem of this theory. By introduction of
background field method this problem has been resolved-although it
could not be applied to the very early moments of evaluation of
universe (the Planck scale) where the perturbation of metric is in
the same order as the background metric itself. The next problem,
which appears in the background field method is the
non-renormaliziability of the theory of quantum gravity. Three
divergent opinions have been stated for explanation of this anomaly.
The first view sees the problem as an inherent problem of general
relativity and the second, as an intrinsic problem of quantum
mechanics. The third group, believe that both, quantum mechanics and
general relativity are problematic and an alternative theory has to
replace them.

Combination of quantum theory and special relativity results in
appearance of singularity in QFT. To cope with this problem
regularization and renormalization procedures, which are not related
to fundamental concepts of quantum mechanic and/or special
relativity, have been successfully utilized to remove divergences
for certain problems. This procedure, however, cannot be extended to
the quantum general relativity, and the theory remains divergent. We
believe the root of this anomaly lies in the QFT, not in general
relativity.

The singular behavior of Green's function at short relative
distances (ultraviolet divergence) or in the large relative
distances (infrared divergence) leads to divergences of the QFT. The
ultraviolet divergence appears in the following terms of Green's
function in the limit $\sigma=0$:
$$\frac{1}{\sigma}, \;\;\ln \sigma, \;\; \mbox{ and } \;\; \delta (\sigma).$$

It was conjectured long ago \cite{des,dew} that quantum metric
fluctuations might smear out the singularities of Green's
functions on the light cone {\it i.e.} $\delta (\sigma)$. Along
the this line the model described by Ford \cite{for2}, which is
based on the quantum metric fluctuations, does smear out the light
cone singularities, but it does not remove other ultraviolet
divergences of quantum field theory.

The quantum field theory in Krein space, {\it i.e.} indefinite
metric quantization, studied previously for other problems
\cite{di,rami}, was utilized for the covariant quantization of the
minimally coupled scalar field in de Sitter space \cite{gareta}. In
this method, the auxiliary negative norm states (negative frequency
solution) have been utilized, the modes of which do not interact
with the physical states or real physical world. One of the
interesting results of this construction is that the quantum field
theory in Krein space removes all ultraviolet divergences of quantum
field theory with exception of the light cone singularity. In the
previous works, we had shown that presence of negative norm states
play the role of an automatic renormalization device for certain
problems \cite{gareta,taijmpe1,taijmpe2,rota,ta2,ta3,knrt}.

In the next section, we have shown the quantum field theory in Krein
space removes all ultraviolet divergences of quantum field theory
with exception of the light cone singularity. In the following
section, we briefly recall that quantum metric fluctuations as a
tool to remove the singularities of Green's functions on the light
cone \cite{for2}. We have established that the combination of
quantum field theory in Krein space together with consideration of
quantum metric fluctuations, results in quantum field theory without
any divergences. Finally, we explicitly calculate for $\lambda
\phi^4$ theory, the transition amplitude of the state
$|q_1,q_2;\mbox{ in}>$ to the state $|p_1,p_2; \mbox{ out}>$ for
s-channel contribution in the one-loop approximation.

\section{Krein space quantization}

Recently, the existence of a non-zero cosmological constant has been
proposed to explain the luminosity observations on the farthest
supernovas \cite{pe,japs}. If this hypothesis is validated, our
ideas on the large-scale universe should to be changed and the de
Sitter metric will play a further important role. Thus the
quantization of the massless tensor spin-$2$ field on dS space, {\it
i.e.} a linear gravitational field, without infrared divergence
presents an excellent modality for further research. The linear
quantum gravity is an important element in the understanding of
quantum cosmology and of quantum gravity. However the graviton
propagator in the linear approximation, for largely separated
points, either has a pathological behavior (infrared divergence) or
violates dS invariance  \cite{altu, flilto,anmo1}. Antoniadis,
Iliopoulos and Tomaras \cite{anilto2} have shown that the
pathological large-distance behavior of the graviton propagator on a
dS background does not manifest itself in the quadratic part of the
effective action in the one-loop approximation. This means that the
pathological behavior of the graviton propagator is gauge dependent
and so should not appear in an effective way as a physical quantity.
Recently, de Vega and al. \cite{vera} have also shown that, the
infrared divergence does not appear in the physical world. This
result has been also obtained by other authors
\cite{hahetu,hiko1,hiko2}. The dS linear gravity could indeed be
constructed from minimally coupled scalar field in the ambient space
notation, {\it i.e.} $k_{\alpha\beta}=D_{\alpha\beta}\phi$
\cite{tathes,gagata,gagarerota}. Thus a large volume of literature
has been devoted to the quantization problem for the
minimally-coupled massless field in de Sitter space \cite{al,kiga}.

It is proven that for the minimally coupled scalar field, in de
Sitter space, one can not construct a covariant quantization with
only positive norm states. In addition there appears an infrared
divergence in the two point function \cite{al}. It has been proved
that the use of the two sets of solutions (positive and negative
norms states) is an unavoidable feature if one is resolved to
preserve causality (locality), covariance and elimination of the
infrared divergence in quantum field theory for the minimally
coupled scalar field in de Sitter space \cite{gareta}. We maintain
the covariance principle and remove the positivity condition similar
to the Gupta-Bleuler quantization of electrodynamics in Minkowski
space.

One of the very interesting results of this construction is that the
Green's function, at large relative distances, does not diverge. In
other words the previous infrared divergence disappears
\cite{gareta,ta3} and the ultraviolet divergence in the stress
tensor disappears as well, which means the quantum free scalar field
in this method is automatically renormalized. The effect of
``unphysical'' states (negative norm states) appears in the physics
as a natural renormalization procedure. By the use of this method
for linear gravity (the traceless rank-2 ``massless'' tensor field)
a fully covariant quantization in dS space is obtained
\cite{gagarerota}. Consequently the corresponding two point function
is free of any infrared divergence \cite{ta2,tathes}. It is
important to note that following this method, a natural
renormalization of the certain problems, have been already attained
\cite{gareta,taijmpe1,taijmpe2,rota,ta2,ta3}.

We briefly recall the Krein space quantization of the minimally
coupled scalar field in de Sitter space \cite{gareta}. As proved by
Allen \cite{al}, the covariant canonical quantization
 procedure with positive norm states fails in this case. The
 Allen's result can be reformulated in the following way: the Hilbert space generated by
 a set of modes (named here the positive modes, including the zero mode) is
 not de Sitter invariant,
 $${\cal H}=\{\sum_{k \geq 0}\alpha_k\phi_k;\;
 \sum_{k \geq 0}|\alpha_k|^2<\infty\}.$$
This means that it is not closed under the action of the de~Sitter
group generators. In order to resolved this problem, we have to deal
with an orthogonal sum of a positive and negative inner product
space, which is closed under an indecomposable representation of the
de~Sitter group. The negative values of the inner product are
precisely produced by the conjugate modes:
$\le\phi_k^*,\phi_k^*\re=-1$, $k\geq 0$. We do insist on the fact
that the space of solution should contain the unphysical states with
negative norm.
 Now, the decomposition of the field operator into positive and
negative norm parts reads
   \b \phi(x)=\frac{1}{\sqrt{2}}\left[ \phi_p(x)+\phi_n(x)\right],\e
 where
 \b \phi_p(x)=\sum_{k\geq 0} a_{k}\phi_{k}(x)+H.C.,\;\;
  \phi_n(x)=\sum_{k \geq 0} b_{k}\phi^*(x)+H.C..\e
 The positive mode $\phi_p(x)$ is the scalar field as was used by Allen.
 The crucial departure from the standard QFT based on CCR lies in the
 following on the commutation relations requirement:
 \b    a_{k}|0>=0,\;\;[a_{k},a^{\dag} _{k'}]= \delta_{kk'},\;\;
   b_{k}|0>=0,\;\;[b_{k},b^{\dag} _{k'}]= -\delta_{kk'}.\e
 A direct consequence of these formulas is the positivity of the energy  {\it i.e.}
$$\le\vec k|T_{00}|\vec k\re\geq0,$$ for any physical state $|\vec
k\re$ (those built from repeated action of the $a^{\dag} _{k}$'s on
the vacuum). This quantity vanishes if and only if $|\vec
k\re=|0\re$. Therefore the ``normal ordering'' procedure for
eliminating the ultraviolet divergence in the vacuum energy, which
appears in the usual QFT is not needed \cite{gareta}. Another
consequence of this formula is a covariant two-point function, which
is free of any infrared divergence \cite{ta3}.

It is important to note that, similar to the gauge quantum field
theory, the auxiliary negative norm states are presented, after the
calculation of the Green function, by imposing the condition on the
field operator, are eliminated. The negative norm states cannot
propagate in the physical world and they only play the rule of an
automatic renormalization device for the theory.

Within the framework of our approach, the ``Wightman'' two-point
function is the imaginary part the usual Wightman two-point
function, which is built from the positive norm states \b {\cal
W}(x,x')=<0\mid \phi(x)\phi(x') \mid 0>=\frac{1}{ 2}[{\cal
W}_p(x,x')+{\cal W}_n(x,x')]=i \Im {\cal W}_p(x,x') ,\e where ${\cal
W}_n=-{\cal W}_p^*$. The time-ordered product of fields is defined
as \b iG_T(x,x')=<0\mid T\phi(x)\phi(x') \mid 0>=\theta (t-t'){\cal
W}(x,x')+\theta (t'-t){\cal W}(x',x).\e In this case we obtain \b
G_T(x,x')=\frac{1}{2}[G_F^p(x,x')+(G_F^p(x,x'))^*]=\Re
G_F^p(x,x').\e For scalar field $\phi(x)$ in $4$-dimensional
Minkowski space-time, the positive norm state time-ordered product
of fields or Feynman two point function is \cite{boloto} $$
G_F^p(x,x')=\int \frac{d^4 k}{(2\pi)^4}e^{-ik.(x-x') }\tilde
G^p(k)=\int \frac{d^4
k}{(2\pi)^4}\frac{e^{-ik.(x-x')}}{k^2-m^2+i\epsilon}$$ \b=
-\frac{1}{8\pi}\delta(\sigma_0)+
 \frac{m^2}{8\pi}\theta(\sigma_0)\left[\frac{J_1(\sqrt {2m^2
 \sigma_0})-iN_1(\sqrt {2m^2 \sigma_0})}{\sqrt {2m^2 \sigma_0}}\right]- \frac{im^2}{4\pi^2}\theta(-\sigma_0)
 \frac{K_1(\sqrt {2m^2(- \sigma_0)})}{\sqrt {2m^2(- \sigma_0)}},\e
where $J_1, N_1$ and $K_1$ are the Bessel functions. Then the
time-ordered product of fields in the Krein space quantization is:\b
G_T(x,x')=\Re G_F^p(x,x')= \frac{-1}{8\pi}\delta(\sigma_0)+
 \frac{m^2}{8\pi}\theta(\sigma_0)\frac{J_1(\sqrt {2m^2
 \sigma_0})}{\sqrt {2m^2 \sigma_0}}, \sigma_0 \geq 0.\e
This function is singular only on the light cone. In the next
section, we briefly recall that quantum metric fluctuations as a
tool to remove the singularities of Green's functions on the light
cone \cite{for2}.

\section{Quantum metric fluctuation}

Reviewing the effective methods of quantum metric fluctuation for
removal of light cone singularities, we present Ford's vivid paper
in this section \cite{for2}. Consideration of a flat background
space time with a linearized perturbation $h_{\mu\nu}$ propagating
upon it, constitute the basic modality of quantum metric
fluctuation, {\it i.e.} \b g_{\mu\nu}=\eta_{\mu\nu}+h_{\mu\nu},\;\;
|h|<|\eta|.\e In the unperturbed space time, the square of the
geodesic separation of points $x$ and $x'$ is $2\sigma_0 =
\eta_{\mu\nu}(x^{\mu}- x'^{\mu})(x^{\nu} - x'^{\nu})$. In the
general curved space-time, the presence of the perturbation
$h_{\mu\nu}$, the square of the geodesic separation is set to be
$2\sigma$, and write \cite{bida} \b \sigma = \sigma_0 + \sigma_1 +
O(h^2),\e so $\sigma_1$ is the first order shift in $\sigma$ and it
is an operator in the linear quantum gravity.

In flat space time, the retarded Green's function for a massless
scalar field is \b G^{(0)}_{ret}(x - x') = \frac{\theta(t - t')
}{4\pi} \delta(\sigma_0),\e which has a delta-function singularity
on the future light cone and is zero elsewhere. In the presence of
a classical metric perturbation, the retarded Green's function has
its delta-function singularity on the perturbed light cone, where
$\sigma = 0$. In general, it may also become nonzero on the
interior of the light cone due to backscattering off of the
curvature. However, we are primarily interested in the behavior of
the near ''new'' light cone, so we replace $G^{(0)}_{ret}(x - x')$
by \b G_{ret}(x - x') = \frac{\theta(t - t') }{4\pi}
\delta(\sigma).\e This may be expressed as \b G_{ret}(x - x') =
\frac{\theta(t - t') }{4\pi} \int_{-\infty}^{\infty}d\alpha
e^{i\alpha\sigma_0}e^{i\alpha\sigma_1}.\e

We now replace the classical metric perturbations by gravitons in a
vacuum state $|\psi\rangle$. Then $\sigma_1$ becomes a quantum
operator which is linear in the graviton field operator,
$h_{\mu\nu}$. Because a vacuum state is a state such that $\sigma_1$
may be decomposed into positive and negative frequency parts, i.e.,
we may find $\sigma^+_1$ and $\sigma^-_1$ so that $\sigma^+_1
|\psi\rangle = 0, \langle\psi\mid \sigma^-_1= 0$, and $\sigma_1
=\sigma^+_1+\sigma^-_1$. Thus when we average over the metric
fluctuations, the retarded Green's function is replaced by its
quantum expectation value: \b \langle G_{ret}(x - x')\rangle =
\frac{\theta(t - t') }{4\pi} \int_{-\infty}^{\infty}d\alpha
e^{i\alpha\sigma_0}e^{i\alpha\langle\sigma_1\rangle}.\e This
integral converges only if $\langle \sigma_1^2 \rangle > 0$, in
which case it may be evaluated to yield \b \langle G_{ret}(x -
x')\rangle = \frac{\theta(t - t') }{8\pi^2}
\sqrt{\frac{\pi}{2\langle\sigma_1^2\rangle}}
exp\left(-\frac{\sigma_0^2}{2\langle\sigma_1^2\rangle}\right).\e
Note that this averaged Green's function is indeed finite at
$\sigma_0 = 0$ provided that $\langle \sigma_1^2 \rangle \neq 0.$
Thus the light cone singularity has been removed.

Therefore the quantum field theory in Krein space, including the
quantum metric fluctuation, removes all ultraviolet divergences of
the theory: \b \langle G_T(x - x')\rangle = -\frac{1 }{8\pi}
\sqrt{\frac{\pi}{2\langle\sigma_1^2\rangle}}
exp\left(-\frac{\sigma_0^2}{2\langle\sigma_1^2\rangle}\right)+
 \frac{m^2}{8\pi}\theta(\sigma_0)\frac{J_1(\sqrt {2m^2
 \sigma_0})}{\sqrt {2m^2 \sigma_0}}.\e In the case of $2\sigma_0 = \eta_{\mu\nu}(x^{\mu}-
x'^{\mu})(x^{\nu} - x'^{\nu})=0$, due to the metric quantum
fluctuation, $h_{\mu\nu}$, $\langle\sigma_1^2\rangle\neq 0$, and we
have \b \langle G_T(0)\rangle = -\frac{1 }{8\pi}
\sqrt{\frac{\pi}{2\langle\sigma_1^2\rangle}} +
 \frac{m^2}{8\pi}\frac{1}{2}.\e
It should be noted that $ \langle\sigma_1^2\rangle $ is related to
the density of gravitons \cite{for2}.

\section{$\lambda
\phi^4$ theory in Krein space}

An immediate consequence of this construction is a finite quantum
field theory. A vivid example is the realization of finite $\lambda
\phi^4$ theory. To demonstrate this we explicitly calculate the
transition amplitude of the state $|q_1,q_2;\mbox{ in}>$  to the
state $|p_1,p_2; \mbox{ out}>$ for s-channel contribution in the
one-loop approximation. It is given by \cite{itzu} $${\cal
T}\equiv<p_1,p_2; \mbox{ out}|q_1,q_2;\mbox{ in}>_s=\int
d^4y_1d^4y_2d^4x_1d^4x_2 e^{ip_1.y_1+ip_2.y_2-iq_1.x_1-iq_2.x_2}$$
$$ (\Box_{y_1}+m^2)(\Box_{y_2}+m^2)(\Box_{x_1}+m^2)(\Box_{x_2}+m^2)
\frac{(-i\lambda)^2}{2!}\int d^4z_1d^4z_2 [iG_T(z_1-z_2)]^2$$ $$
G_T(y_1-z_2)G_T(y_2-z_1)G_T(x_1-z_2)G_T(x_2-z_2),$$ where the
Feynman Green function $G_F^p$ is replaced by the time-order product
Green function $G_T$. We obtain $$ {\cal T}=\frac{\lambda^2}{2}\int
d^4z_1d^4z_2 e^{i(p_1+p_2).z_1-i(q_1+q_2).z_2}[G_T(z_1-z_2)]^2
=\frac{\lambda^2}{2}(2\pi)^4 \delta^4(p_1+p_2-q_1-q_2)$$ \b \int
d^4z e^{i(p_1+p_2).z}\left( -\frac{1 }{8\pi}
\sqrt{\frac{\pi}{2\langle\sigma_1^2\rangle}}
exp\left(-\frac{\sigma_0^2}{2\langle\sigma_1^2\rangle}\right)+
 \frac{m^2}{8\pi}\theta(\sigma_0)\frac{J_1(\sqrt {2m^2
 \sigma_0})}{\sqrt {2m^2 \sigma_0}}\right)^2,\e where $2\sigma_0=(z_1-z_2)^2=z^2$. The seconde integral for
the space-like separated pair $(z_1,z_2)$ is zero. The first
integral is gaussian and the seconde is a fourier transformation of
a non singular function. Therefore the transition amplitude in Krein
space quantization, is finite in the one-loop approximation.

\section{Conclusion}

The problem of divergence disappears in QFT, if the principle of the
positivity of the norms is overlooked. In the other word, the
problem appears when the negative norm states, which are also
solutions of the field equation in the classical level, are
eliminated by one of the principles of quantum theory, i.e. the
principle of positivity. The direct result of the principle of
positivity is appearance of distractive anomalies in QFT,
divergences and breaking of the covariance of the minimally coupled
scalar field and linear quantum gravity in de Sitter space.

Implementing of this method to the linear gravity (traceless part)
on de Sitter space, a covariant two-point function, free of the
pathological large-distance behavior was obtained \cite{ta2}. The
quantum field theory in Krein space, which includes quantum metric
fluctuations, resolves the ultraviolet divergences of quantum field
theory. Finally the theory of quantum gravity in the back ground
field method is achieved without any divergencies. \vspace{0.5cm}

\noindent {\bf{Acknowlegements}}: The authors would like to thank T.
Banks for helpful discussion.

\end{document}